\documentclass[a4paper,twoside,twocolumn,english,aps,twocolumn,floatfix,showpacs,tightenlines,superscriptaddress,groupedaddress]{revtex4}
\usepackage{times}
\usepackage[]{fontenc}
\usepackage[latin1]{inputenc}
\usepackage{amsmath}
\usepackage{color}
\usepackage{graphicx}
\usepackage{amssymb}

\makeatletter

\newcommand{\lyxdot}{.}

\usepackage{babel}
\makeatother
\begin{document}

\title{Rapidly-converging methods for the location of quantum critical points
from finite-size data}

\author{M. Roncaglia}

\affiliation{Max-Planck-Institut f\"ur Quantenoptik, Hans-Kopfermann-Str. 1,
D-85748, Garching, Germany}

\author{L. Campos Venuti}

\affiliation{Fondazione ISI, Villa Gualino, viale Settimio Severo 65, I-10133,
Torino, Italy}

\author{C. Degli Esposti Boschi}

\affiliation{CNR, Unità di Ricerca CNISM and Dipartimento di Fisica dell'Università
di Bologna, viale Berti-Pichat 6/2, I-40127, Bologna, Italy}

\begin{abstract}
We analyze in detail, beyond the usual scaling hypothesis, the finite-size
convergence of static quantities toward the thermodynamic limit. In
this way we are able to obtain sequences of pseudo-critical points
which display a faster convergence rate as compared to currently used
methods. The approaches are valid in any spatial dimension and for
any value of the dynamic exponent. We demonstrate the effectiveness
of our methods both analytically on the basis of the one dimensional
XY model, and numerically considering $c=1$ transitions occurring
in non integrable spin models. In particular, we show that these general
methods are able to locate precisely the onset of the Berezinskii-Kosterlitz-Thouless
transition making only use of ground-state properties on relatively
small systems. 
\end{abstract}


\maketitle

\section{Introduction}

In the study of physical properties of phase transitions a basic prerequisite
is a reliable method to locate the critical point, whenever the latter
is not known \textit{a priori} from symmetry or duality arguments.
Typically, in numerical or even experimental studies on finite samples
one obtains a sequence of pseudocritical points (in the sense specified
below) to be extrapolated to the true critical point in the thermodynamic
limit (TL). The extrapolation may be done with some polynomial fit
in the inverse size of the system or, better, exploiting some fitting
function derived on the basis of a scaling \textit{ansatz} or through
the renormalization group (RG). The point is especially relevant in
the context of quantum phase transitions (QPT) \cite{sachdev99} in
lattice systems where the exponential growth of the dimension of the
Hilbert space with the number of sites is a strong limitation on the
accessible sizes with the current computational power and algorithms.
One of the most used algorithms is still the Lanczos method for the
virtually exact extraction of the low-lying energy levels; in the
most favorable case of spin-1/2 models one cannot go beyond some tens
of sites. This limit can be moved to maybe a few thousands of sites
using the so-called density matrix renormalization group (DMRG) \cite{schollwock05}
that has become the method of choice for 1D problems due to its high
level of accuracy. Nonetheless, if one considers two or even three
dimensional systems the situation is much worse: with the Lanczos
algorithm the largest lattices have only a few sites of linear extension
and the DMRG is not particularly efficient. At present, the only other
choice is Quantum Monte Carlo (QMC) (see, for instance, \cite{evertz03})
that, however, suffers from a sign problem in the case of fermionic
or frustrated systems and does not reach the level of accuracy of
the DMRG. Very recently there have been attempts to exploit both DMRG-like
features and the QMC sampling tricks, to design hybrid methods \cite{sandvik07,schuch07}
that are however still under verification.

It is generally believed that a sequence of pseudocritical points,
for example the loci of maxima of finite-size susceptibilities, converges
to the critical point as a power law $L^{-\lambda}$, with a so-called
\textit{shift} exponent $\lambda$ given by the inverse of the correlation
length exponent $\nu$. Hence, generally speaking, the larger is $\nu$
the slower is the convergence. This difficulty reaches its maximum
for Berezinskii-Kosterlitz-Thouless (BKT) transitions, in which the
correlation length diverges with an essential singularity or, loosely
speaking, {}``$\nu=\infty$''. However, already in the seminal paper
by Fisher and Barber \cite{fisher82}, it was pointed out that the
relation $\lambda=1/\nu$ is not always valid and $\lambda$ depends,
among other factors, on the boundary conditions. 

The most used method to locate quantum critical points in $d=1$ by
means of finite-size data is the so-called Phenomenological Renormalization
Group (PRG), reviewed for instance in \cite{barber83}. Another convenient
approach, the Finite-size Crossing Method (FSCM), was recently proposed
in \cite{campos_local_06}. The aim of this paper is to improve both
of them by means of criteria that produce sequences of pseudocritical
points that converge more rapidly. We will show that in our sequences
the shift exponent will have the form $\lambda=\lambda_{0}+\lambda_{1}/\nu$,
where $\lambda_{0}$ and/or $\lambda_{1}$ are larger than the corresponding
values in the usual methods and therefore allow for a better convergence. 

The paper is organized as follows. In Section \ref{sec:rcs} we illustrate
the general arguments leading to the enhanced sequences, both in the
framework of the FSCM and of the PRG (Subsec.~\ref{sub:PRG}). Special
cases as the BKT transition (Subsec.~\ref{sub:BKT}) and that of
logarithmic divergences (Subsec.~\ref{sub:log}) are discussed separately.
In Section \ref{sec:test} we illustrate the usefulness of the methods
on the hand of analytic and numerical tests. In Subsec.~\ref{sub:XY}
we treat the XY spin-1/2 chain, using a series of exact calculations
reported in the Appendix. Then we move to two cases of spin chains
for which no exact solution is available: in Subsec.~\ref{sub:JzD}
we consider a spin-1 model with anisotropies in a parameter range
that gives rise to a large value of $\nu$ and in Subsec.~\ref{sub:J1J2}
we study the spin-1/2 model with next to nearest neighbor interactions
that is known to undergo a BKT transition. In this case we find a
value for the critical coupling in agreement with the accepted one,
which was found using a model-specific investigation of the excited
states \cite{okamoto92}. Section \ref{sec:conc} is devoted to conclusions.

\section{Derivation of rapidly-converging sequences\label{sec:rcs}}

We consider systems in $d$ spatial dimensions of linear size $L$
and periodic boundary conditions (PBC). Let the transition be driven
by a linear parameter $g$ such that the Hamiltonian is 

\[
\mathcal{H}(g)=\mathcal{H}_{0}+g\mathcal{V}.\]

Dealing with QPT we consider the case of strictly zero temperature,
$T=0$, even if the arguments presented below can be simply extended
to the finite-temperature case, replacing the parameter $g$ with
$T$ (and without using the \textit{dimensional crossover rule} used
below). The free energy density reduces to the ground-state (GS) energy
density which, close to the critical point $g_{c}$, shows a singularity
in the second (or higher) derivatives with respect to $g$: 

\[
\frac{1}{L^{d}}\langle\mathcal{H}(g)\rangle=e(g)=e_{\mathrm{reg}}(g)+e_{\mathrm{sing}}(\xi(g)),\]
where $\xi\approx|g-g_{c}|^{-\nu}\equiv t^{-\nu}$ is the correlation
length. Note that, as a consequence of the scaling hypothesis, the
singular part of the energy $e_{{\rm sing}}$ is a universal quantity
that depends only on $\xi$, the relevant length scale close to the
critical point. Hence, $e_{{\rm sing}}$ may be considered quite in
general an even function of $(g-g_{c})$ that vanishes at the critical
point. 

On the other hand, the bulk energy density at the critical point behaves
as (Privman-Fisher hypothesis) \begin{equation}
e(g_{c},L)=e_{\infty}(g_{c})-L^{-(d+\zeta)}F(g_{c}),\label{eq:PFh}\end{equation}
where $F(g)$ is a sort of Casimir-like term that may depend on the
actual geometry of the lattice. Note that this hypothesis has to be
changed properly if one or more of the spatial dimensions are of infinite
extent. Moreover, Eq.~(\ref{eq:PFh}) has been written in analogy
with Eq.~(11.29) of ref.~\cite{brankov00} using the dimensional
crossover rule according to which the partition function and the thermodynamic
(static) properties of a $d$-dimensional quantum system are equivalent
to those of a $(d+\zeta$)-dimensional classical counterpart \cite{sachdev99,brankov00},
where $\zeta$ is the dynamic exponent \cite{sachdev99}. Then, for
the implementation of our methods we need to know by some other means
the value of $\zeta$ relating the energy gap $\Delta$ and the correlation
length $\xi$: $\Delta\propto\xi^{-\zeta}$. Typically, but not always,
energy and momentum in the continuum limit at the critical point satisfy
a \textit{linear} dispersion relation, $E=vk$, for small $k$ so
that $\zeta=1$ and a relativistic effective field theory can be used
to describe the universal features of the transition. In $d=1$ the
scale invariance at the critical point is often sufficient to imply
also conformal invariance (see Ch. 2 of \cite{henkel99}), thanks
to which several exact results can be obtained using the powerful
predictions of conformal field theories (CFT). For example, by mapping
the space-time complex plane onto a cylinder whose circumference represents
the finite chain of length $L$ we can identify $F(g_{c})=\pi cv(g_{c})/6$
in Eq.~(\ref{eq:PFh}) where $c$ is the central charge of the theory.
In the RG sense, moving away from criticality corresponds to perturbing
the CFT with a relevant operator that destroys conformal invariance.
However, this is not the only effect of varying the microscopic parameter
$g$ out of $g_{c}$: in general also the speed of elementary excitations
gets renormalized in the unperturbed CFT part. For this reason we
say that $v$ (henceforth $F$) depends on $g$, in the vicinity of
$g_{c}$. 

Scaling and dimensional arguments imply that, in the thermodynamic,
off-critical regime $L\gg\xi$, the singular part of the energy behaves
as\begin{equation}
e_{\mathrm{sing}}\approx t^{2-\alpha}\label{eq:e_sing}\end{equation}
with $\alpha=2-\left(d+\zeta\right)\nu$. For a second order phase
transition $\alpha<1$.

After introducing the scaling variable $z=(L/\xi)^{1/\nu}=tL^{1/\nu}$,
the finite-size scaling (FSS) theory asserts \cite{barber83} that
in a system of length $L$,

\begin{equation}
e_{\mathrm{sing}}=C_{0}L^{-(d+\zeta)}\Phi_{e}(z)+\cdots,\label{eq:e_fss}\end{equation}
where $\Phi_{e}(z)$ is a universal function that, in the off-critical
regime $z\gg1$, must behave as $\Phi_{e}(z)\approx z^{(d+\zeta)\nu}$
in order to recover Eq.~(\ref{eq:e_sing}) . Instead, for $L\ll\xi$
we are in the critical regime and $\Phi_{e}(z)$ behaves as an analytic
function that vanishes for $z\to0$. Here we assume that the leading
term in $\Phi_{e}(z)$ is quadratic in $z$ (see below why it cannot
be linear), but the following arguments are easily generalizable to
higher integer powers. The constant term $C_{0}\Phi_{e}(0)/L^{d+\zeta}$
is already adsorbed in the nonuniversal part of the energy density
at $g_{c}$ as shown in Eq. (\ref{eq:PFh}).

Differentiating $e(g)$ with respect to $g$, gives the mean value
$b=\langle\mathcal{V}\rangle/L^{d}$, whose singular part $b_{\mathrm{sing}}$
behaves as\begin{equation}
b_{\mathrm{sing}}\approx\textrm{sgn}(g-g_{c})t^{1-\alpha}=\textrm{sgn}(g-g_{c})t^{(d+\zeta)\nu-1}.\label{eq:b_sing}\end{equation}
Considering FSS for the combination of Eqs.~(\ref{eq:PFh}) and (\ref{eq:e_fss})
and then differentiating we find\begin{multline}
b(g,L)=b_{\infty,{\rm reg}}(g)+\textrm{sgn}(g-g_{c})C_{0}L^{1/\nu-(d+\zeta)}\Phi_{e}^{\prime}(z)\\
-L^{-(d+\zeta)}\left[F^{\prime}(g_{c})+F^{\prime\prime}(g_{c})(g-g_{c})+\cdots\right]+O\left(L^{-(d+\zeta+\epsilon)}\right)\label{eq:b_fss}\end{multline}
where the subscript {}``$\infty$, reg'' hereafter means regular
in the TL. In order to write down the expression above we used $\partial_{g}=L^{1/\nu}\partial_{z}$
and assumed that the powers neglected in the last term are just larger
than $(d+\zeta)$. To illustrate this point we could consider the
irrelevant operator with the smallest scaling dimension $d_{{\rm irr}}$.
At first order in perturbation theory with the renormalized coupling
$g_{{\rm irr}}(L)=g_{{\rm irr}}(0)L^{d+\zeta-d_{{\rm irr}}}$ the
corrections to the GS energy density are of the form $C_{{\rm irr}}g_{{\rm irr}}(0)L^{-d_{{\rm irr}}}$
so that $\epsilon=d_{{\rm {\rm irr}}}-d-\zeta>0$. Note also that
the amplitude $C_{{\rm irr}}$ can vanish and $O(g_{{\rm irr}}^{2})$
terms have to be included. For these and more details we leave the
reader to Ref.~\cite{Cardy86}. Generically we admit corrections
with $\epsilon\ge0$ that may come either from irrelevant operator
in the continuum theory or from lattice effects. The case $\epsilon=0$
corresponds to marginal perturbations and typically leads to logarithmic
corrections. Notice that now the leading term of $\Phi_{e}^{\prime}(z)$
is linear in $z$ in the critical region. If we would have admitted
a linear term in $\Phi_{e}(z)$ then $\Phi_{e}^{\prime}(0)\neq0$
and a finite jump discontinuity at finite $L$ would be present in
$b(g,L)$. 

We can also calculate $a=\langle\mathcal{H}_{0}\rangle/L^{d}=e(g)-g[\partial e(g)/\partial g]$,
yielding to a singular part that is similar to Eq.~(\ref{eq:b_sing})
but with a changed sign. In fact, for $g_{c}\not=0$ the leading singular
parts of $a$ and $b$ must cancel in the sum that gives back the
energy $e(g)=a(g)+gb(g)$, which does not contain that singularity.
In particular, the scaling is\begin{eqnarray*}
a(g,L) & = & a_{\infty,{\rm reg}}(g)-\textrm{sgn}(g-g_{c})gC_{0}L^{1/\nu-(d+\zeta)}\Phi_{e}^{\prime}(z)\\
 &  & +C_{0}L^{-(d+\zeta)}\Phi_{e}(z)\\
 &  & -L^{-(d+\zeta)}\left[F(g)-gF^{\prime}(g_{c})+O(g-g_{c})\right]+\dots\end{eqnarray*}
The FSCM \cite{campos_local_06} identifies the critical point with
the limit of the sequence $g_{L}^{*}$ of single crossing points \[
b\left(g_{L}^{*},L\right)=b(g_{L}^{*},L^{\prime})\]
with $L^{\prime}=L+\delta L$. Applying this criterion to Eq.~(\ref{eq:b_fss}),
we obtain (for $\delta L\ll L$)\[
g_{L}^{*}-g_{c}=-\frac{(d+\zeta)F^{\prime}(g_{c})}{{2C}_{0}\left[\frac{2}{\nu}-(d+\zeta)\right]}L^{-2/\nu}.\qquad\nu\not=\frac{2}{d+\zeta}\]
This equation defines the shift exponent $\lambda_{{\rm FSCM}}=2/\nu$
and may converge very slowly when $\nu\gg1$. The extremely difficult
case is the BKT transition where formally $\nu=\infty$, but this
latter situation must be treated in a different way (see Subsec. \ref{sub:BKT}). 

Now, consider the quantity $\Gamma(g,L,\gamma)=\gamma e(g,L)-b(g,L)$
and suppose to be able to tune $\gamma$ exactly at \begin{eqnarray}
\gamma^{*} & = & \frac{F^{\prime}(g_{c})}{F(g_{c})}.\label{eq:balance}\end{eqnarray}
It easily seen that $\Gamma(g_{c},L,\gamma^{*})$ does not contain
the Casimir-like term responsible for the critical point shift. In
fact, the scaling of $\Gamma$ is\begin{multline*}
\Gamma(g,L,\gamma)=\Gamma_{\infty,{\rm reg}}(g,\gamma)-\textrm{sgn}(g-g_{c})C_{0}L^{1/\nu-(d+\zeta)}\Phi_{e}^{\prime}(z)\\
-L^{-(d+\zeta)}\left[\gamma F(g)-F^{\prime}(g_{c})+O(g-g_{c})\right]+O\left(L^{-(d+\zeta+\epsilon)}\right).\end{multline*}
When $\gamma$ is equal to $\gamma^{*}$ given in Eq.~(\ref{eq:balance}),
the critical point found by this crossing method is approached as
$g_{L}^{\ast}-g_{c}\approx L^{-\lambda_{{\rm fast}}}$, with a shift
exponent $\lambda_{{\rm fast}}=2/\nu+\epsilon=\lambda_{{\rm FSCM}}+\epsilon$
(again this holds true provided that $\nu\neq2/(d+\zeta)$). The additional
term $\epsilon$ allows, in general, for a better convergence of the
sequence $g_{L}^{*}$.

A possible algorithm for finding numerically the critical point in
such a way is the following. If $g_{L}^{*}$ is at a crossing point,
of $\Gamma(g,L,\gamma)$, then we have \[
\gamma(g_{L}^{*},L)=\frac{b(g_{L}^{*},L+\delta L)-b(g_{L}^{*},L)}{e(g_{L}^{*},L+\delta L)-e(g_{L}^{*},L)}.\]
Notice that putting $\gamma=0$ is equivalent to the FSCM applied
to $b(g,L)$ and the denominator has definite sign about the critical
point. Now we find $g_{L}^{*}$ requiring that $\gamma(g_{L}^{*},L-\delta L)=\gamma(g_{L}^{*},L)$,
i.e. \begin{equation}
\frac{b(g_{L}^{*},L)-b(g_{L}^{*},L-\delta L)}{e(g_{L}^{*},L)-e(g_{L}^{*},L-\delta L)}=\frac{b(g_{L}^{*},L+\delta L)-b(g_{L}^{*},L)}{e(g_{L}^{*},L+\delta L)-e(g_{L}^{*},L)},\label{eq:crossing_fast_fd}\end{equation}
or in the continuum version \begin{equation}
\frac{\partial_{L}b(g,L)}{\partial_{L}e(g,L)}=\frac{\partial_{L}^{2}b(g,L)}{\partial_{L}^{2}e(g,L)}\label{eq:crossing_fast}\end{equation}

Calling $\tau=(g-g_{c})=t\mathrm{sgn}(g-g_{c})$ and rewriting only
the essential terms in the scaling \textit{ans}{\it \"a}\textit{tze}
we have the simplified forms \begin{eqnarray*}
e(g,L) & = & e_{\infty,{\rm reg}}(g)-L^{-(d+\zeta)}F(g)+D_{1}L^{-(d+\zeta+\epsilon)},\\
b(g,L) & = & b_{\infty,{\rm reg}}(g)+L^{-(d+\zeta)}\left[{2C}_{0}L^{2/\nu}\tau-F^{\prime}(g)\right]\\
 &  & +D_{2}L^{-(d+\zeta+\epsilon)}.\end{eqnarray*}
Now, putting these two relations in (\ref{eq:crossing_fast}) we obtain 

\begin{equation}
g_{L}^{*}-g_{c}=\frac{D_{2}F(g_{c})-D_{1}F^{\prime}(g_{c})}{4C_{0}F(g_{c})}\:\frac{\nu\epsilon(d+\zeta+\epsilon)}{(d+\zeta-\frac{2}{\nu})}\: L^{-\lambda_{{\rm fast}}}\label{eq:gLfast}\end{equation}
that gives a shift exponent $\lambda_{\mathrm{fast}}$ as anticipated.

The main result of this section is the crossing criterion (\ref{eq:crossing_fast_fd})
that identifies the rapidly converging sequence (\ref{eq:gLfast})
to the critical value.

\subsection{Homogeneity condition}

In the previous section we have shown how to obtain a sequence of
pseudocritical points with an improved shift exponent $\lambda_{{\rm fast}}$.
Here we provide another, yet simpler equation for the determination
of the critical point. The resulting pseudocritical sequence is characterized
by the same shift exponent $\lambda_{{\rm fast}}$ . However in this
case we are able to prove convergence toward the critical point even
in the extreme case of a BKT transition (see Subsec.~\ref{sub:BKT}).

The idea is to require that at the critical point the $L$-dependent
part of $b$ is dominated by the Casimir-like term with power $(d+\zeta)$
(see Eq.~(\ref{eq:b_fss})). This condition is translated into the
requirement that the $L$-derivative of $b$ is a homogeneous function
of degree $(d+\zeta+1)$, i.e. \begin{equation}
(d+\zeta+1)\partial_{L}b(g_{L}^{\ast},L)+L\partial_{L}^{2}b(g_{L}^{\ast},L)=0.\label{eq:hcgc}\end{equation}
Consequently, the corresponding sequence of pseudocritical points
$g_{L}^{*}$ scales as\begin{equation}
g_{L}^{*}-g_{c}=\frac{\epsilon D_{2}(d+\zeta+\epsilon)}{{2C}_{0}\frac{2}{\nu}(d+\zeta-\frac{2}{\nu})}L^{-\lambda_{{\rm fast}}}.\label{eq:gLhom}\end{equation}

The equation (\ref{eq:hcgc}) represents the \emph{homogeneity condition
method} (HCM) that we are proposing for the efficient location of
critical points. We stress here that, being $b\left(g,L\right)$ a
GS property, this criterion does not require knowledge of excited
states as in the case of the PRG method. This is a point in favor
to the HCM since excited states are typically assessed with less numerical
accuracy. In addition, the HCM is superior to the PRG in that it produces
a faster converging sequence (see Subsec.~\ref{sub:PRG}).

\subsection{Case $\nu=2/(d+\zeta)$ with logarithmic divergences\label{sub:log}}

For completeness we consider the case $\nu=\frac{2}{d+\zeta}$ that
was excluded in the previous treatment. In this situation, the \textit{ansatz}
requires the inclusion of logarithmic corrections \[
b(g,L)=b_{\infty,{\rm reg}}(g)+C_{1}\ln L\tau-L^{-(d+\zeta)}F^{\prime}(g)+D_{2}L^{-(d+\zeta+\epsilon)}.\]
 The calculation of the critical point with the FSCM gives in this
case

\[
g_{L}^{*}-g_{c}=-\frac{(d+\zeta)F^{\prime}(g_{c})}{C_{1}}L^{-(d+\zeta)}\]
whereas the HCM Eq.~(\ref{eq:hcgc}) yields\[
g_{L}^{*}-g_{c}=-\frac{\epsilon D_{2}(d+\zeta+\epsilon)}{C_{1}(d+\zeta)}L^{-(d+\zeta+\epsilon)}\]
These results are compatible with the exact calculations of the XY
model (see Subsec. \ref{sub:XY}). Note that in general, we should
to perform a similar calculation for \[
\nu=\frac{p}{d+\zeta},\qquad p\in\mathbb{N}+1,\]
 namely when the $p$-th derivative of the free energy diverges logarithmically.

\subsection{A scaling \textit{ansatz} for the BKT case\label{sub:BKT}}

For $d=1$ at the BKT the correlation function in the TL behaves like
$\xi\approx\exp(at^{-\sigma})$. In the typical example of the classical
two dimensional XY model it is known that $\sigma=1/2$. Instead,
for the quantum Heisenberg model with frustration (which we will consider
in Subsec. \ref{sub:J1J2}) Haldane suggested $\sigma=1$ \cite{okamoto92}.
We also set $\zeta=1$ because the (effective) dimensionality in the
BKT scenario is two. The singular part of the finite-size energy density
now is conveniently expressed in terms of $y\equiv L/\xi$ \cite{barber83}
so that \[
e(g,L)=e_{\infty,{\rm reg}}(g)+L^{-2}\left[{C_{0}\Phi}_{e}(y)-F(g)\right]+O\left(L^{-(2+\epsilon)}\right)\]
where $\Phi_{e}(y)$ is a universal function that, in the off-critical
regime $y\gg1$, must behave as $\Phi_{e}(y)\approx y^{2}$. Again,
in the quasicritical regime $y\ll1$ at any finite $L$ the energy
density and its derivatives must be analytic in $(g-g_{c})$. The
value $\Phi_{e}(0)$ can be absorbed in $F(g)$ and it can be checked
directly that the first contribution has to be at least quadratic
in $t=\vert g-g_{c}\vert$ because otherwise a finite-size discontinuity
in $b(g,L)$ would be generated. For $y\ll1$ we adopt the following
\textit{ansatz} (justified from perturbed conformal field theory \cite{Cardy86}):\[
e(g,L)=e_{\infty,{\rm reg}}(g)+L^{-2}\left[K(at^{-\sigma}-\ln L)^{-n/\sigma}-F(g)\right]+\dots\]
with $K$ a constant and $n$ an integer larger than 1 ($n=3$ from
Eq (22) in \cite{Cardy86}). Hence\begin{multline}
b(g,L)=b_{\infty,{\rm reg}}(g)\\
+L^{-2}\left[nKa^{-n/\sigma}t^{n-1}\mathrm{sgn}(g-g_{c})\times\right.\\
\left.\left(1-\frac{\ln L}{a}t^{\sigma}\right)^{-\frac{n}{\sigma}-1}-F'(g)\right]+O\left(L^{-(2+\epsilon)}\right)\label{eq:b_BKT}\end{multline}
Now we want to get rid of all the $O(L^{-2})$ contributions that
{}``hinder'' the location of finite-size pseudocritical points.
Hence we first differentiate with respect to $L$ to eliminate $b_{\infty,{\rm reg}}(g)$,
then multiply by $L^{3}$ to isolate the term in square brackets in
Eq.(\ref{eq:b_BKT}) and finally set to zero a further difference
in $L$ in order to drop $F'(g)$. Formally, in the region $L/\xi\ll1$,
we can write down the condition \begin{equation}
\partial_{L}\left[L^{3}\partial_{L}b(g_{L}^{\ast},L)\right]=0.\label{eq:hcgcBKT}\end{equation}
It is worth noticing that the latter condition is equal to the HCM
Eq.~(\ref{eq:hcgc}) when $d=\zeta=1$. Treating $L$ as a continuous
variable one can read off the shift exponent for the sequence $g_{L}^{*}$
that turns out to be $\lambda_{{\rm fast}}^{{\rm BKT}}=\epsilon/(n-1+\sigma)$;
if $\epsilon=0$ then the corrections to scaling are also governed
by (another) marginal operator and we expect $g_{L}^{*}-g_{c}\sim(\ln L)^{-(m+1)/(n-1+\sigma)}$
with $m$ a positive integer ($m=4$ from Eq.~(22) in \cite{Cardy86}).

Dealing with numerical simulations it is very important to specify
how one implements the finite-size differences in $L$. In fact, there
are several finite-difference expressions used in the literature to
express the derivatives and here the requirement is that they all
reproduce Eq.~(\ref{eq:hcgcBKT}) in the limit $L\to\infty$. For
example, if one takes a uniform step $\delta L$ then the following
symmetric expression can be built\begin{eqnarray*}
b'(g,L) & \equiv & \frac{b(g,L+\delta L)-b(g,L-\delta L)}{2\delta L},\\
b''(g,L) & \equiv & \frac{b(g,L+\delta L)-2b(g,L)+b(g,L-\delta L)}{(\delta L)^{2}},\end{eqnarray*}
and then the precise condition to cancel the term $O(L^{-2})$ becomes\begin{equation}
L^{3}b''(g_{L}^{*},L)+\left[3L^{2}-(\delta L)^{2}\right]b'(g_{L}^{*},L)=0.\label{eq:hcgcfL}\end{equation}
In the limit of large $L$ we recover (\ref{eq:hcgcBKT}) as required.
Clearly, the correct discretization prescription must be identified
not only for the BKT; when $d$ and/or $\zeta$ are not one, the suitable
variant of (\ref{eq:hcgcfL}) for finite $L$ has to be adopted, with
$b''$ weighted by $L^{(d+\zeta+1)}$ and $b'$ weighted by a polynomial
in $L$ of degree $(d+\zeta)$ whose coefficients depend on $(\delta L)$.

\subsection{PRG revisited\label{sub:PRG}}

The PRG method identifies the critical point with the limit of the
sequence $g_{L}^{*}$ of crossing points satisfying\begin{equation}
L^{\zeta}\Delta\left(g_{L}^{*},L\right)-(L+\delta L)^{\zeta}\Delta(g_{L}^{*},L+\delta L)=0.\label{eq:prg}\end{equation}
Here $\Delta(g,L)$ is the finite-size energy gap of the spectrum
for which we may adopt the following form \begin{equation}
\Delta(g,L)=L^{d}\left[e_{1}(g,L)-e(g,L)\right]=L^{-\zeta}\Phi(z)+L^{-(\zeta+\epsilon)}C_{\chi}\label{eq:ansatz_delta}\end{equation}
with $\Phi(z)\simeq\Phi_{0}+z\Phi_{1}+z^{2}\Phi_{2}+\dots$ and $C_{\chi}$
a prefactor depending on the excited state $\vert\chi\rangle$ we
are considering. The standard PRG approach in $d=1$ relies upon
Eq.~(\ref{eq:prg}) with $\zeta=1$. More generally, a first test
to identify $\zeta$ is done by plotting the usual scaled gaps $L\Delta$
and see if at the critical point they settle to a constant or not.
If they do not, one is led to search for a better value of $\zeta\neq1$
and solve (\ref{eq:prg}) with the correct value of the dynamic exponent.
Note that the corrections come from the irrelevant/lattice contributions
and the Casimir-like term for the gap is actually the constant term
in $\Phi(z)$. For quantum systems in $d=1$, CFT ($\zeta=1$) predicts
$\Phi_{0}=2\pi d_{\chi}v(g)$, where $d_{\chi}$ is the scaling dimension
of the operator $\hat{\chi}$ that generates the excited state $\vert\chi\rangle$.
In addition $\Phi_{1},\Phi_{2},\dots$ can be computed in the framework
of perturbed CFT, in the regime $z\ll1$ when a relevant operator
$(g-g_{c}){\cal R}$ with scaling dimension $x=2-1/\nu$ is added
to the critical field theory. For instance, from Eqs.~(7) and (10)
of Ref.~\cite{Cardy86} we have $\Phi_{1}=b_{\chi}/(2\pi)^{x}$ where
$b_{\chi}$ is the structure constant that appears as prefactor of
the three points correlation function $\langle{\cal R}(\vec{r}_{1})\hat{\chi}(\vec{r}_{2})\hat{\chi}(\vec{r}_{3})\rangle$. 

Now, in general the PRG method gives $\vert g_{L}^{*}-g_{c}\vert\approx L^{-\lambda_{{\rm PRG}}}$
with $\lambda_{{\rm PRG}}=1/\nu+\epsilon$ \cite{henkel99}. However,
it may happen that $\Phi_{1}=0$ in the scaling function (examples
are discussed in Subsec. \ref{sub:JzD} and at the end of Subsec.~\ref{sub:XY}
for a specific excited state of the XY model). The shift exponent
in such a case decreases to $\lambda_{{\rm PRG}}=1/\nu+\epsilon/2$.
Nonetheless there is a way to improve this behavior. In fact, the
{}``extremum'' (instead of the zeroes) of the quantity in left side
of Eq.(\ref{eq:prg}) with the \textit{ansatz} (\ref{eq:ansatz_delta})
is located exactly at the critical point $z=0$. In order to appreciate
a shift from criticality we have to include higher orders of $\tau=(g-g_{c})$
in the irrelevant non-scaling term, i.e. \[
L^{\zeta}\Delta(g,L)=\Phi_{0}+L^{2/\nu}\tau^{2}\Phi_{2}+\left(C_{\chi}(0)+C_{\chi}'(0)\tau\right)L^{-\epsilon}.\]
In this case, the convergence is $\vert g_{L}^{*}-g_{c}\vert\approx L^{-\lambda_{{\rm extr}}}$
with $\lambda_{{\rm extr}}=2/\nu+\epsilon$, which is better than
usual PRG not only because of the double exponent but also thanks
to the coefficient in front: it is proportional to $C_{\chi}'(0)$
which is usually a small quantity. It is also worth noticing that
in this case $\lambda_{{\rm {\rm extr}}}=\lambda_{{\rm fast}}$.

\section{Testing the methods\label{sec:test}}

\subsection{XY model in transverse field\label{sub:XY}}

To check our methods analytically, we consider the 1D spin-1/2 XY
model given by\begin{equation}
H=-\sum_{j=1}^{L}\left[\frac{(1+\eta)}{2}\sigma_{j}^{x}\sigma_{j+1}^{x}+\frac{(1-\eta)}{2}\sigma_{j}^{y}\sigma_{j+1}^{y}+h\sigma_{j}^{z}\right].\label{eq:HXY}\end{equation}
Throughout the paper we will consider $L$ even and PBC. This model
can be solved exactly \cite{lsm61,henkel99} by means of a Jordan-Wigner
transformation from spins to spinless fermions followed by a Bogoliubov
transformation to arrive at a Hamiltonian of free quasiparticles.
The number of original fermions $\mathcal{N}=\sum_{i}c_{i}^{\dagger}c_{i}$,
is not a conserved quantity, but its parity $P\equiv\exp(i\pi\mathcal{N})$
corresponds to a $\pi$-rotation around the $z$-axis, and therefore
is conserved. One should beware of a delicate issue concerning boundary
conditions. Starting with PBC in Eq.~(\ref{eq:HXY}), the fermionic
Hamiltonian turns to have PBC in the sector of odd parity $P=-1$.
Instead in the even parity sector $P=1$ -- which comprises the ground
state -- antiperiodic boundary conditions must be used. In this sector
the model becomes\begin{equation}
H=\sum_{k}\Lambda\left(k\right)\left(\beta_{k}^{\dagger}\beta_{k}-\frac{1}{2}\right),\label{eq:XY_quasip}\end{equation}
where $\beta_{k}$ are Bogoliubov quasiparticles, $k$ ranges in the
first Brillouin zone and the dispersion relation is\begin{equation}
\Lambda\left(k\right)=2\sqrt{\eta^{2}+h^{2}+(1-\eta^{2})\cos^{2}k+2h\cos k}\,.\label{eq:drXY}\end{equation}
For $\eta\neq0$ this model displays an Ising transition at $h=1$
with exponents $\nu=\zeta=1$, which means that the scaling variable
is $z=L\left(h-1\right)$. To test our ideas we need to calculate,
for finite $L$, the GS energy $e_{L}\left(h,\eta\right)$ and the
average potential $b_{L}\left(h,\eta\right)=-\langle\sigma_{j}^{z}\rangle=\partial e_{L}/\partial h$,
in the quasi-critical region given by $z\ll1$. According to Eq.~\eqref{eq:XY_quasip}
the GS is given by $e_{L}\left(h,\eta\right)=-1/2L\sum_{k_{n}}\Lambda\left(k_{n}\right)$
with $k_{n}=(2n+1)\pi/L,\; q=0,\dots,L-1$. We then expand the argument
of the sum up to the desired order in $z$. For our purposes we need
$e_{L}(h,\eta)$ up to $O(z^{2})$ and $b_{L}\left(h,\eta\right)$
up to $O\left(z\right)$. The resulting sums are then evaluated with
the aid of Euler-Maclaurin formula \cite{atkinson_book}. The results
and the details are given in the Appendix \ref{sec:app}. From Eqs.~\eqref{eq:e_XY}
and \eqref{eq:b_XY}, one sees explicitly that terms of order $L^{-3}$
are absent both from $e_{L}$ and $b_{L}$ so that according to our
definition Eq.~\eqref{eq:b_fss} we find $\epsilon=2$. In passing
we notice that, from Eq.~(\ref{eq:e_XY}), the Casimir-like term
is ${-L}^{-2}\pi\vert\eta\vert/6$ consistent with the CFT formula
$-L^{-2}\pi cv/6$. In fact the central charge is $c=1/2$ and, from
the dispersion relation (\ref{eq:drXY}) the spin velocity turns out
to be $v=2\vert\eta\vert$. 

Now, using Eqs.~\eqref{eq:e_XY} and \eqref{eq:b_XY}, we have all
the elements to derive the sequences of pseudocritical points analytically.
As far as the FSCM is concerned, the pseudo critical points are obtained
imposing $b_{L}\left(h_{L}^{\ast}\right)=b_{L+2}\left(h_{L}^{\ast}\right)$,
or formally $\partial_{L}b_{L}\left(h_{L}^{\ast}\right)=0$. Up to
leading order in $L$ the solution is \[
h_{L}^{*}=1+\frac{\pi^{2}}{6L^{2}}\,,\]
as already obtained in Ref.~\cite{campos_local_06} for the Ising
model $(\vert\eta\vert=1)$. This shows explicitly that $\lambda_{{\rm FSCM}}=2$
consistent with our prediction $\lambda_{{\rm FSCM}}=2/\nu$ (remind
that $\nu=1$). 

For what concerns the {}``balancing trick'' discussed in Sec.~\ref{sec:rcs},
we can show that the solution of Eq.~(\ref{eq:balance}), is given
by $\gamma^{*}=-1/2\eta^{2}$. The pseudocritical points are then
given imposing $\partial_{L}\Gamma(h_{L}^{\ast},L,\gamma^{*})=0$.
In this case, at leading order, the solution is \[
h_{L}^{*}=1-\frac{7\pi^{4}(2\eta^{4}+\eta^{2}-3)}{720\eta^{2}}L^{-4},\]
which means $\lambda_{{\rm fast}}=4$ once again consistent with our
prediction $\lambda_{{\rm fast}}=2/\nu+\epsilon$ (remind that $\epsilon=2$).
From a numerical point of view it is more profitable to use the HCM.
From $3\partial_{L}b+L\partial_{L}^{2}b=0$ we obtain\[
h_{L}^{\ast}=1+\frac{7\pi^{4}\left(2\eta^{2}-3\right)}{720\eta^{2}}L^{-4}\]
with the same exponent $\lambda_{{\rm fast}}=4$. 

Now we proceed to discuss the PRG method for which knowledge of the
lowest gap is required. The first excited state belongs to the sector
with odd parity. Correspondingly, the finite size gap is given, besides
a constant term, by the difference between two Riemann sums where
the sampling is taken over odd and even (in units of $\pi/L$) wavenumbers
respectively. In analogy with Ref.~\cite{hamer1981} the lowest gap
can be eventually written as $\Delta_{L}=2(h-1)+[T(L)-2T(L/2)]$,
where $T(L)=\left(1/2\right)\sum_{j=0}^{2L-1}\Lambda(j\pi/L)$. Again
we refer to the Appendix for the details. Using the form of the gap
Eq.~\eqref{eq:XY_gap} we can evaluate the various terms of the scaling
function in Eq.~\eqref{eq:ansatz_delta}\begin{equation}
\Phi(z)=\frac{\pi\vert\eta\vert}{2}+z+\frac{\pi^{3}\left[3\left(h-1\right)-2\left(2+h\right)\eta^{2}+8\eta^{4}\right]}{192\left|\eta\right|^{3}}L^{-2}+\dots\label{eq:PhiPRG}\end{equation}
Hence $\Phi_{0}=\pi\vert\eta\vert/2=2\pi vd_{\chi}$, consistently
with Ref.~\cite{hamer1981} for $\vert\eta\vert=1$ and with the
known result $d_{\chi}=1/8$ for the first excited state of the $c=1/2$
minimal model \cite{henkel99}. Moreover, since $\Phi_{1}\neq0$,
according to the general analysis in Subsec.~\ref{sub:PRG}, there
is no advantage in searching the minima or maxima in $\tau$ of the
left hand side of Eq.~(\ref{eq:prg}). Looking for the zeros by imposing
the PRG equation $\partial_{L}\left(L\Delta_{L}\left(h_{L}^{\ast}\right)\right)=0$,
yields \[
h_{L}^{*}=1+\frac{\pi^{3}\left(4\eta^{2}-3\right)}{48\left|\eta\right|}L^{-3}+O\left(L^{-5}\right).\]
This result for generic $\eta\neq0$, shows explicitly that the PRG
shift exponent is $\lambda_{{\rm PRG}}=3$ and proves a conjecture
put forward by Hamer and Barber in Ref.~\cite{hamer1981} for the
Ising model. From Eq.~(\ref{eq:PhiPRG}) together with Eq. \eqref{eq:ansatz_delta}
we read $\epsilon=2$ as previously, so that the calculated exponent
is consistent with the prediction $\lambda_{{\rm PRG}}=1/\nu+\epsilon$. 

The explicit calculation of the various shift exponents, confirms
that the HCM given by Eq.~(\ref{eq:hcgc}) is superior to the standard
PRG method. However, the things are different if we considered the
excitation obtained creating a well-defined quasiparticle with the
smallest momentum $k=2\pi/L$, instead of the very first excitation
gap. In that case, the energy gap is given by Eq.~(\ref{eq:drXY}).
Using the method of the extrema of the PRG quantity in the left hand
side of Eq.~(\ref{eq:prg}), we would have obtained a convergence
to the critical point as $L^{-4}$, while with the standard PRG only
as $L^{-2}$.

\subsection{$c=1$ transition (non BKT)\label{sub:JzD}}

We choose the spin-1 $J_{z}-D$ model on a chain ($d=1$) with PBC\begin{equation}
H=J\sum_{j=1}^{L}\vec{S}_{j}\cdot\vec{S}_{j+1}+(J_{z}-J)S_{j}^{z}S_{j+1}^{z}+D(S_{j}^{z})^{2}\label{eq:HJzD}\end{equation}
because the transition from the Haldane phase to the phase at large
$D$ is described by a $c=1$ CFT ($\zeta=1$) with continuously varying
exponents. The Hamiltonian (\ref{eq:HJzD}) has been used to describe
the magnetic properties of different quasi 1D compounds (see \cite{cristian04}
for a brief account). Here, by fixing $J=1$ and $J_{z}=0.5$, for
which it has been already estimated $\nu=2.38$ \cite{cristian03},
we wish to test the methods described above in a severe case in which
a $1/\nu$ scaling would give \textit{\textcolor{black}{sublinear}}
convergence in $L$. Indeed for the FSCM we expect $\lambda_{{\rm FSCM}}=0.84$.
The previous estimate using the log-log plots of the finite-size gaps
was $D_{c}=0.65$ \cite{cristian03}, while in Ref.~\cite{chen03}
it is found $D_{c}=0.635$ using the method of level crossing with
antiperiodic boundary conditions, which is however specific to this
transition. The first irrelevant operator allowed by the lattice symmetries
has scaling dimension $K+2$ where $K=2-\frac{1}{\nu}$ so that $\epsilon=\min(K,\epsilon_{{\rm lattice}})$
where with $\epsilon_{{\rm lattice}}$ we denote the smallest exponent
of corrections arising from lattice contributions at finite $L$ that
are not captured in the framework of the (relativistic) continuum
theory.

As discussed in Refs.~\cite{kitazawa97,cristian04} this is a case
in which the linear term $\Phi_{1}$ in the scaling function of the
PRG vanishes because $b_{\chi}=0$ for the sine-Gordon model, the
effective field theory that describe the surroundings of the $c=1$
line. So it is convenient to use also our improved version of the
PRG as discussed in Subsec. \ref{sub:PRG}. The expected shift exponent
is $\lambda_{{\rm extr}}=2/\nu+\epsilon$.

In order to have an idea of the range of values of $L$ to be used
let us imagine that $\epsilon=K$ so that $\lambda_{{\rm fast}}=\lambda_{{\rm extr}}=2+1/\nu=2.42$.
With $L>10$ this exponent leads to variations in the pseudocritical
points $D_{L}^{*}$ smaller than $O(10^{-3})$. Hence we prefer to
illustrate the method with virtually exact numerical data obtained
with the Lanczos algorithm using $L=8,10,12,14,16$. The reason is
that the DMRG would give rather accurate values for the energies but
the estimates for $b=\langle(S^{z})^{2}\rangle$ could not be sufficiently
precise to appreciate the variations in $D_{L}$ obtained from the
crossings. So we use the DMRG only to extend the data to $L=18,20$
with $3^{7}$ optimized states. In any case the GS belongs to the
$S_{{\rm tot}}^{z}=0$ sector. 

\begin{figure}
\includegraphics[width=6.5cm,clip]{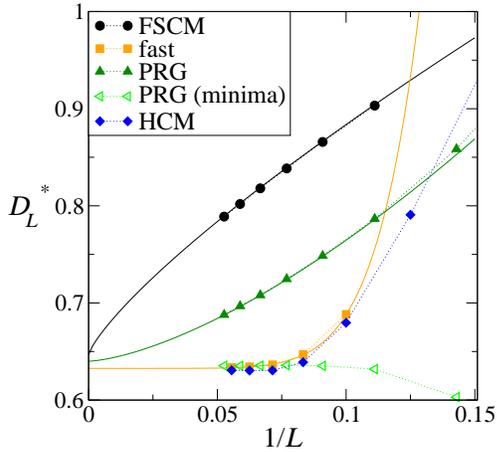}

\caption{(color online) The sequences of pseudocritical points obtained with
the FSCM, the fast-convergent method of Eq.~(\ref{eq:crossing_fast_fd})
and the HCM described by Eq.~(\ref{eq:hcgcfL}). Moreover the sequences
have been calculated with PRG methods, both standard and in our improved
version. Continuous lines are algebraic best-fits to the data.\label{fig:FSCM+PRG+hom}}
\end{figure}

The FSCM with $b=\langle{(S}_{j}^{z})^{2}\rangle$ yields $D_{c}^{{\rm FSCM}}=0.647$
with $\lambda_{{\rm FSCM}}=0.79$ in reasonable agreement with the
CFT expectations. The high-precision procedure based on Eq.~(\ref{eq:crossing_fast_fd})
yields $D_{c}^{{\rm fast}}=0.633$ with $\lambda_{{\rm fast}}=7.6$.
The value of the shift exponent is definitely larger than what expected,
which could be due to the vanishing of the coefficient of the first
irrelevant contribution with scaling dimensions $K+2<\epsilon_{{\rm lattice}}$.
In any case, from Fig. \ref{fig:FSCM+PRG+hom} one can clearly appreciate
that the sequence $D_{L}^{{\rm FSCM}}$ converges more slowly than
$D_{L}^{{\rm fast}}$. As far as the PRG is concerned, with the standard
procedure of finding the zeroes of Eq.~(\ref{eq:prg}) we get $D_{c}^{{\rm PRG}}=0.640$
with $\lambda_{{\rm PRG}}=1.50$. Instead, using the improved method
estimation by looking for the extremal value of the quantity on the
left side of (\ref{eq:prg}) we find a rapidly converging sequence
that however also oscillates between $0.636$ and $0.635$. Finally,
a small oscillation in the fourth decimal place about $0.6305$ is
seen also in the sequence $D_{L}^{{\rm hom}}$ obtained through the
homogeneity criterion (\ref{eq:hcgcfL}). As expected the latter sequence
converges in a fashion similar to $D_{L}^{{\rm fast}}$. All these
results are summarized in Fig. \ref{fig:FSCM+PRG+hom}.

The data show that subtracting out the terms that induce a slow convergence
of the pseudocritical points, one remains with quantities that inherit
a residual (possibly oscillating) $L$-dependence from the specific
lattice model and that could be very difficult to account for. Other
factors that affect the extrapolation of the critical points at this
level of accuracy are the sampling $\delta D$ and the trade-off between
computational accuracy and the maximum available size. 

To summarize we observe that our improved methods yield very fast
convergence to the critical point that we estimate to be $D_{{\rm c}}=0.633\pm0.02$,
consistently with Ref.~\cite{chen03}.

\subsection{$c=1$ BKT transition\label{sub:J1J2}}

\begin{figure}
\includegraphics[width=6.5cm,keepaspectratio,clip]{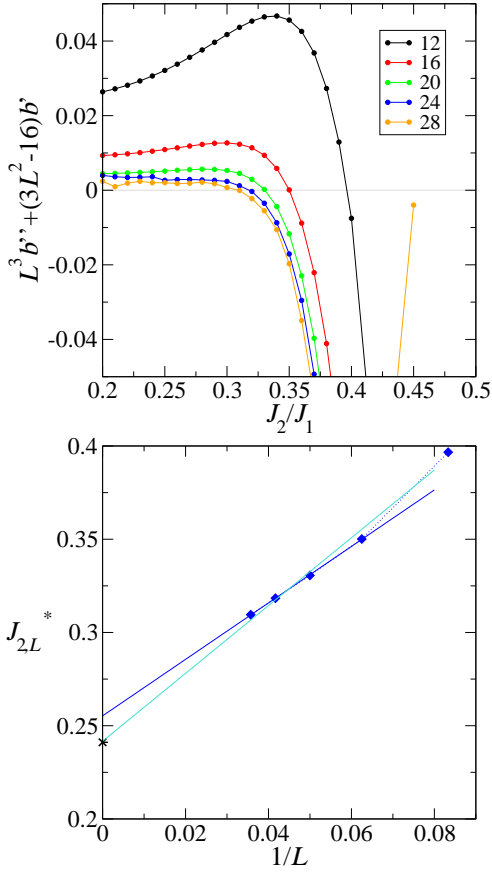}

\caption{\noindent (color online ) Upper panel: searching for the zeroes of
$L^{3}b''+(3L^{2}-16)b'$ according to Eq.~(\ref{eq:hcgcfL}) with
$J_{1}=1$ and $\delta L=4$. Lower panel: extrapolations of the pseudocritical
points $J_{2,L}$ (cyan line for all the points, blue line with $L=12$
excluded); the black cross marks the accepted value $0.2411$ (see
text for further details). \label{fig:homBKT}}
\end{figure}

We consider now the spin-1/2 Heisenberg model with frustration due
to next-to-nearest neighbors interaction\[
H=\sum_{j=1}^{L}J_{1}\vec{S_{j}}\cdot\vec{S}_{j+1}+J_{2}\vec{S_{j}}\cdot\vec{S}_{j+2}.\]
The model is equivalent to a 2-legs zigzag ladder with $L/2$ rungs.
The best estimate of the critical point was made by Okamoto and Nomura
using a model-specific crossing method \cite{okamoto92}; with exact
diagonalizations up to $L=24$ they determined $J_{2{\rm c}}/J_{1}=0.2411\pm0.0001$.
The model is gapless for $J_{2}<J_{2{\rm c}}$ and has a doubly degenerate
GS in the TL for $J_{2}>J_{2{\rm c}}$. Again the GS has $S_{{\rm tot}}^{z}=0$.
Without exploiting \textit{\textcolor{black}{a priori}} information
about the BKT character of transition (if not the value of the dynamic
exponent $\zeta=1$), we tested the homogeneity criterion (\ref{eq:hcgcfL})
using ladders with PBC of up to $L/2=16$ rungs, an effective $\delta L=4$
and $1024$ DMRG states that ensure an accuracy of $O(10^{-7})$ on
the values of $b=\langle\vec{S}_{j}\cdot\vec{S}_{j+2}\rangle$. The
results reported in Fig. \ref{fig:homBKT} are encouraging: while
with the FSCM we would get no crossings at all, the zeroes of Eq.~(\ref{eq:hcgcfL})
yield a sequence of points converging to $J_{2}\sim0.25$. The main
problem comes from the left side of the transition where the truncation
DMRG error for $L=28$ and $32$ induces some oscillations on the
plotted quantity. We content ourselves with linear fits in $1/L$.
If we exclude the point with $L=12$ the fit is better even if we
find $J_{2{\rm c}}/J_{1}=0.2553\pm0.0008$; by selecting all the available
points, instead, the fit is visibly worse but the extrapolated value
is $J_{2{\rm c}}/J_{1}=0.242\pm0.006$, in agreement with Ref.~\cite{okamoto92}.
As above, apart from the details of the extrapolation procedure, we
see that the homogeneity criterion provides a viable procedure to
locate the critical point in a BKT transition, where almost all existing
generic methods fail. We remark that this analysis is based solely
on $b(g,L)$, namely an observable evaluated on the GS, without invoking
further assumptions on the nature of the excitations. 

\section{Conclusions\label{sec:conc}}

Making only use of finite-size quantities related to the ground state,
we show how to generate sequences of pseudocritical points that converge
very fast to the infinite-size critical point. The convergence is
of the form $L^{-\lambda}$ with a shift exponent $\lambda$. In this
article we propose a \emph{homogeneity condition method} (HCM) which
is faster than the standard \emph{phenomenological renormalization
group} (PRG) in locating the critical points. Moreover its validity
is more general as it can be applied without modification to the difficult
case of a Berezinskii-Kosterlitz-Thouless transition. The homogeneity
method requires only the knowledge of $b(g,L)=\langle{\cal V}\rangle/L^{d}$,
that is, the expectation value of the term that drives the transition. 

We also presented an improvement to the PRG method, which allows,
under certain conditions, to obtain pseudocritical sequences characterized
by the same shift exponent as for the HCM. However, this modification,
relying on a particular form of the gap scaling function, is not valid
in general. It holds true, for instance for the sine-Gordon model
that underlies a variety of transitions in $\left(1+1\right)$ dimension.

The formulations of the approaches are sufficiently general to be
applied in any spatial dimensionality. Even if we are primarily interested
in quantum phase transitions ideally at zero temperature, in principle
the methods can be extended to problems of finite-temperature statistical
mechanics. At variance with other accelerating methods found in the
literature, e.g.~the van den Broeck-Schwartz or the Bulirsch-Stoer
ones (reviewed in Ch.~9 of \cite{henkel99}), the procedures presented
here rely on the scaling behavior of thermodynamical quantities expected
from physical and renormalization arguments.

The validity of the methods has been tested with analytical calculations
on the one dimensional XY model in transverse field and numerically
on a nontrivial spin-1 chain with anisotropy. As extreme case, we
have shown that the homogeneity condition method provides a satisfactory
location of the critical point also in the case of Berezinskii-Kosterlitz-Thouless
transitions. These confirmations motivate us to consider systems in
higher spatial dimensionality. In these cases the numerical data are
restricted to smaller system-size and the need for fast-converging
pseudo-critical sequences is a prerequisite for the precise location
of the critical points. 

\begin{acknowledgments}
This work was partially supported by the Italian MiUR through the
PRIN grant n. 2005021773. M.R. acknowledges support from the EU (SCALA).
Numerical calculations were performed on a cluster of machines made
available by the Theoretical Group of the Bologna Section of the INFN.
\end{acknowledgments}
\begin{appendix}

\section{\label{sec:app}}

Here we indicate how to compute the mean energy per site $e_{L}$,
and the average potential $b_{L}$, as require id in Subsec.~\ref{sub:XY}.
Consider for example the energy sum\[
e_{L}(h,\eta)=-\frac{1}{2L}\sum_{n=0}^{L-1}\Lambda\left(k_{n}\right),\quad k_{n}=\frac{2n+1}{L}\pi,\]
where $\Lambda(k)$ is given by Eq.~(\ref{eq:drXY}). We need to
investigate the above sum in the quasi-critical region $z=L/\xi\ll1$
so it is sufficient to expand $e_{L}(h,\eta)$ in powers of $(h-1)=z/L$\begin{eqnarray*}
e_{L}(h,\eta) & = & -\frac{1}{2L}\sum_{n=0}^{L-1}\left[\Lambda_{h=1}\left(k_{n}\right)+\frac{z}{L}\partial_{h}\Lambda\vert_{h=1}\left(k_{n}\right)\right.\\
 &  & \left.+\left(\frac{z}{L}\right)^{2}\frac{1}{2}\partial_{h}^{2}\Lambda\vert_{h=1}\left(k_{n}\right)+O\left(z^{3}\right)\right].\end{eqnarray*}
The resulting sums can be computed using the Euler-Maclaurin formula
(see e.g.~\cite{atkinson_book})\begin{multline}
\delta\sum_{j=0}^{N}f(a+j\delta)=\int_{a}^{b}f(x){\rm d}x+\frac{\delta}{2}\left[f(a)+f(b)\right]+\\
\sum_{r=0}^{m}\frac{\delta^{2r}}{(2r)!}B_{2r}\left[f^{(2r-1)}(b)-f^{(2r-1)}(a)\right]+R_{m}\label{eq:eul-mac}\end{multline}
valid for a function $f$ with at least $2m$ continuous derivatives
in $(a,b)$. Here $B_{n}$ are the Bernoulli numbers and the remainder
$R_{m}$ depends on $f^{\left(2m\right)}$ on $\left(a,b\right)$.
Some care must be taken when the function $f$ diverges at the border
of the Brillouin zone, in this case one must keep $a$ and $b$ away
from the borders. Moreover, sending $m$ to infinity in Eq.~\eqref{eq:eul-mac},
some sums must be regularized using a Borel summation technique. The
final result for the ground state energy is\begin{align}
e_{L}\left(h,\eta\right) & =e_{\infty}\left(1,\eta\right)-\frac{\pi\left|\eta\right|}{6}L^{-2}-\frac{7\pi^{3}}{360}\frac{\left(4\eta^{2}-3\right)}{4\left|\eta\right|}L^{-4}\nonumber \\
 & +\left(h-1\right)\left[b_{\infty}\left(1,\eta\right)-\frac{\pi}{12\left|\eta\right|}L^{-2}\right]\nonumber \\
 & -\frac{\left(h-1\right)^{2}}{2}\frac{\ln\left(L\right)+\gamma_{C}+\ln\left(8\left|\eta\right|/\pi\right)-1}{\pi\left|\eta\right|}+O\left(z^{3}\right)\label{eq:e_XY}\end{align}
where $\gamma_{C}=0.577216\dots$ is the Euler-Mascheroni constant
and the thermodynamic values are given by \begin{eqnarray*}
e_{\infty}\left(1,\eta\right) & = & -\left[\frac{2\left|\eta\right|}{\pi}+\frac{2\arctan\left(\sqrt{1-\eta^{2}}/\left|\eta\right|\right)}{\pi\sqrt{1-\eta^{2}}}\right]\\
b_{\infty}\left(1,\eta\right) & = & -\frac{2\arctan\left(\sqrt{1-\eta^{2}}/\left|\eta\right|\right)}{\pi\sqrt{1-\eta^{2}}}.\end{eqnarray*}
Using similar procedures we obtain the following expression for average
potential $b_{L}\left(h,\eta\right)$:\begin{align}
b_{L}\left(h,\eta\right) & =b_{\infty}\left(1,\eta\right)-\frac{\pi}{12\left|\eta\right|}L^{-2}-\frac{7\pi^{3}(3-2\gamma^{2})}{2880\left|\eta\right|^{3}}L^{-4}\nonumber \\
 & -\left(h-1\right)\frac{\ln\left(L\right)+\gamma_{C}+\ln\left(8\left|\eta\right|/\pi\right)-1}{\pi\left|\eta\right|}+O\left(z^{2}\right).\label{eq:b_XY}\end{align}
Finally, the sum\[
T(L)=\frac{1}{L}\sum_{j=0}^{L-1}\Lambda\left(\frac{2\pi j}{L}\right)\]
for the evaluation of the finite-size gap in the PRG method, can be
treated along similar lines. The final result for the gap is\begin{eqnarray}
\Delta_{L} & = & \left(h-1\right)+\frac{\pi\left(2\eta^{2}+h-1\right)}{4\left|\eta\right|}L^{-1}\label{eq:XY_gap}\\
 & + & \frac{\pi^{3}\left[3\left(h-1\right)-2\left(2+h\right)\eta^{2}+8\eta^{4}\right]}{192\left|\eta\right|^{3}}L^{-3}+O\left(L^{-5}\right).\nonumber \end{eqnarray}

\end{appendix}

\bibliographystyle{apsrev}
\bibliography{ref}

\end{document}